\documentstyle[aps,multicol,psfig]{revtex}

\begin{document}
\draft

\title{Reducing vortex density in superconductors using the ratchet effect}
\author{C.-S. Lee$^1$, B. Jank\'o$^2$, I. Der\'enyi$^3$, and A.-L. Barab\'asi$^1$}
\address{$^1$Department of Physics, University of Notre Dame,
         Notre Dame, IN 46556  \\
	$^2$Materials Science Division, Argonne National Laboratory, 9700 South Cass Avenue, Argonne, IL 60439 USA\\
        $^3$Dept. of Surgery, MC 6035, University of Chicago,
            5841 S. Maryland Ave., Chicago, IL 60637}
\date{\today}

\maketitle


\pacs{PACS numbers: }

\begin {multicols}{2}
\narrowtext


{\bf A serious obstacle that impedes the application of low
  and high temperature superconductor (SC) devices is the
  presence of trapped flux \cite{tinkham,trapped}. Flux lines or vortices are
  induced by fields as small as the Earth's magnetic field. Once present, vortices dissipate energy and
  generate internal noise, limiting the operation of numerous
  superconducting devices \cite{trapped,ruggiero}.  Methods used to
  overcome this difficulty include the pinning of vortices by the
  incorporation of impurities and defects \cite{rpm}, the construction
  of flux dams\cite{koch}, slots and holes \cite{dantsker2} and
  magnetic shields \cite{trapped,ruggiero} which block the penetration
  of new flux lines in the bulk of the SC or reduce the magnetic field
  in the immediate vicinity of the superconducting device.  Naturally,
  the most desirable would be to remove the vortices from the bulk of the
  SC. There is no known phenomenon, however, that could form the basis
  for such a process. Here we show that the application of an ac
  current to a SC that is patterned with an asymmetric pinning
  potential can induce vortex motion whose direction is determined
  only by the asymmetry of the pattern.  The mechanism responsible for this
  phenomenon is the so called ratchet
  effect\cite{magnasco,astumian,julicher}, and its working principle
  applies to both low and high temperature SCs. As a first step here we demonstrate that with an appropriate
  choice of the pinning potential the ratchet
  effect can be used to remove vortices from low temperature SCs in the parameter range
  required for various applications.}

\begin{figure}[htb]
\begin{center}
\mbox{\psfig{figure=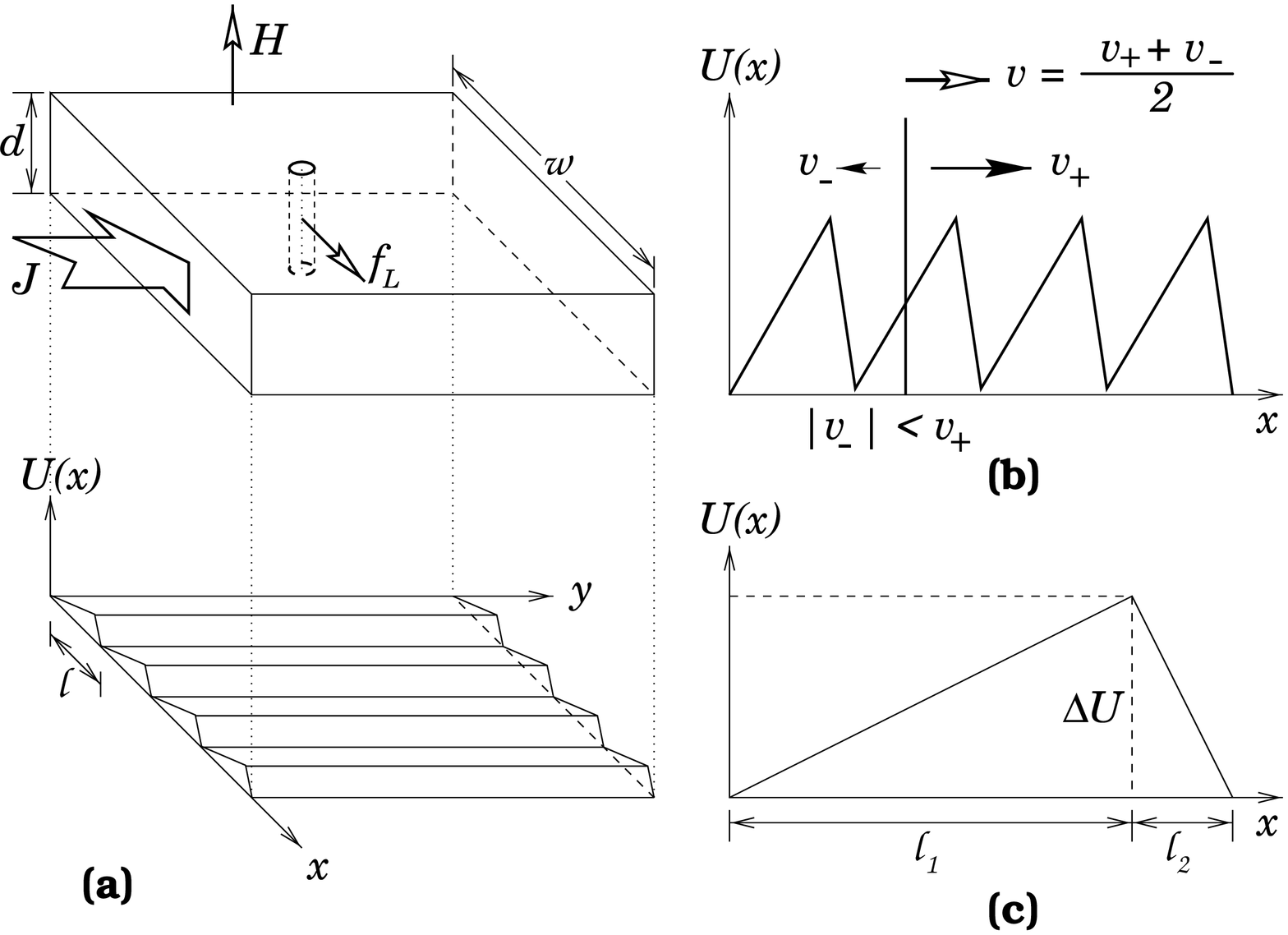,width=12 cm,height=8 cm,angle=-90}}
\end{center}
\protect \caption{ Patterning with an asymmetric potential of a SC.
  {\bf (a)} Schematic  illustration of a SC in the presence of an external
  magnetic field $H$. A dc current with density $J$ flowing along the
  $y$ direction (indicated by the
  large arrow) induces a Lorentz force $f_{\rm L}$ that moves the vortex in 
  the $x$ direction. The SC is patterned with a pinning
  potential $U(x,y)=U(x)$, whose shape  is shown in the lower panel. The
  potential is periodic and asymmetric along the $x$ direction, and is 
  translationally invariant along $y$. {\bf (b)} The pinning potential
  $U(x)$ along the SC cross section. The solid arrows  indicate the
  vortex velocity $v_+$ ($v_-$) induced by a direct $+J$
  (reversed $-J$) current. The average,  $v = (v_+ + v_-)/2$, is the
  ratchet velocity of the vortex, obtained when an ac current is
  applied. {\bf (c)} The parameters  characterizing a single tooth of the
  asymmetric potential.}
\label{Fig1}
\end{figure}

Consider a type II superconductor film of the geometry shown in Fig.\ 
\ref{Fig1}, placed in an external magnetic field $H$.  The
superconductor is patterned with a pinning potential $U(x,y)=U(x)$ which is
periodic with period $\ell$ along the $x$ direction, has an asymmetric
shape within one period, and is translationally invariant along the
$y$ direction of the sample. The simplest example of an asymmetric
periodic potential, obtained for example by varying the sample thickness, is the asymmetric sawtooth potential, shown in Fig.\ 
\ref{Fig1}b. In the presence of a current with density
${\bf J}$ flowing along the $y$ axis the vortices move with the
velocity
\begin{equation}
{\bf v} = ({\bf f_{\rm L}} + {\bf f_{vv}} + {\bf f_u})/\eta,
\label{vel0}
\end{equation}
where \({\bf f_{\rm L}}=({\bf J}\times{\bf \hat h}) \Phi_0 d/c\) is the Lorentz
force moving the vortices transverse to the current, ${\bf \hat h}$ is the
unit vector pointing in the direction of the external magnetic field
${\bf H}$, \({\bf f_u} = -{{\rm d} U \over {\rm d} x} {\bf \hat{x}}\) is
the force generated by the periodic potential, \({\bf f_{vv}}\) is the
repulsive vortex-vortex interaction, $\Phi_0 = 2.07\times10^{-7}$~G$\,$cm$^2$ is the flux quantum, \(\eta\) is the viscous drag
coefficient, and $d$ is the length of the vortices (i.e. the thickness of the sample).

When a dc current flows along the positive $y$ direction, the
Lorentz force moves the vortices along the positive $x$ direction with
velocity $v_+$. Reversing the current reverses the direction of the vortex velocity, but its magnitude, $|v_-|$, due to the asymmetry of the potential, is different from $v_+$. For the
sawtooth potential shown in Fig.\ \ref{Fig1}b the vortex velocity is
higher when the vortex is driven to the right, than when it is driven to
the left ($v_{+}>|v_{-}|$). As a consequence the application of an
ac current (which is the consecutive application of direct and
reverse currents with density $J$) results in a net velocity $v = (v_+ +
v_-)/2$ to the right in Fig.\ \ref{Fig1}b. This net velocity induced by
the combination of an asymmetric potential and an ac driving force is
called ratchet velocity \cite{magnasco,astumian,julicher,hanggi}.  The ratchet velocity for the case of
low vortex density (when vortex-vortex interactions are neglected) can
be calculated analytically. Denoting the period of the ac current by $T$,
the ratchet velocity of the vortices in the $T\to \infty$ limit is given by
the expression
\begin{equation}
v = \left\{ \begin{array}{ll}
        0 & \mbox{if $f_{\rm L}<f_1$} \\
        {1 \over 2\eta}{(f_{\rm L}+f_2)(f_{\rm L}-f_1) \over f_{\rm L}+f_2-f_1} & \mbox{if $f_1<f_{\rm L}<f_2$} \\
        {1 \over \eta}{f_1f_2(f_2-f_1) \over f_{\rm L}^2-(f_2-f_1)^2} & \mbox{if $f_2<f_{\rm L}$},
        \end{array}
        \right.
\label{vel1}
\end{equation}
where $f_1=\Delta U / \ell_1$ and $f_2=\Delta U / \ell_2$
are the magnitudes of the forces generated by the ratchet potential on the facets of length 
$\ell_1$ and $\ell_2$, respectively (see Fig.\ \ref{Fig1}c),
$\Delta U$ is the energy difference between the maximum and the minimum of the potential,
and $f_{\rm L} = |{\bf f_{\rm L}}| = J\Phi_0 d/c$.

\begin{figure}[htb]
\begin{center}
\vskip -3.0 cm
\hskip -2.0 cm
\mbox{\psfig{figure=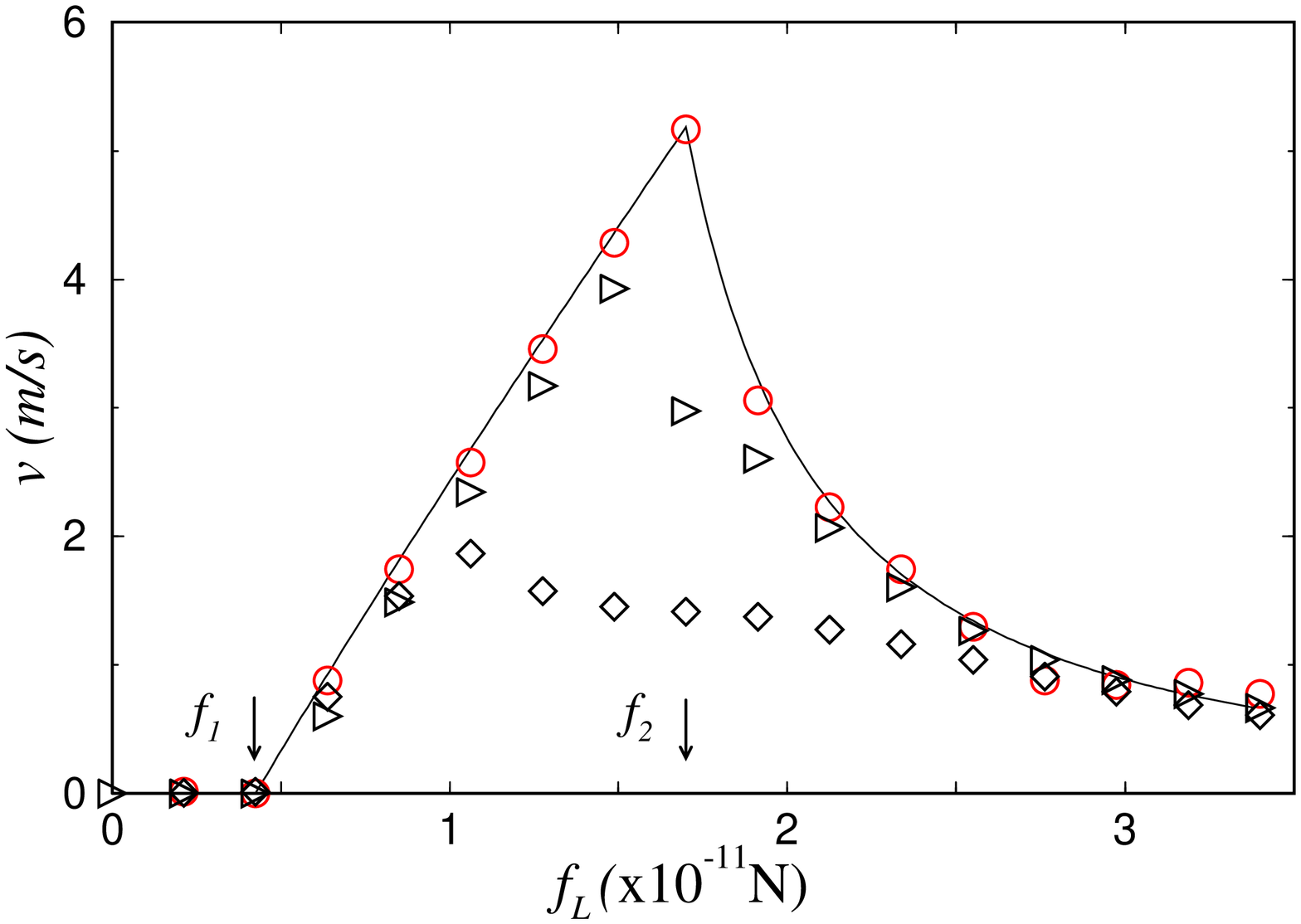,height=8 cm,angle=-90}}
\vskip 0.1 cm
\end{center}
\protect \caption { 
  Ratchet velocity of the vortices as a function of the amplitude of the driving force $f_{\rm L}$. The thick
  solid line corresponds to the analytical result (\protect \ref{vel1}) for a
  single vortex line. The symbols are the result of the numerical
  simulations for multiple vortices. The simulations were done using
  the model developed by Nori and collaborators \protect \cite{Nori,reichhardt78,olson98}, assuming
  that the rigid vortices are pointlike object moving in the $x$-$y$
  plane. At time zero the vortices are positioned randomly in the SC
  with a density $\rho_0$ and they move with velocity given by
  (\protect \ref{vel0}).  The vortex-vortex interaction between two vortices at
  position ${\bf r}_i$ and ${\bf r}_j$ is modeled using \protect \cite{tinkham} 
  ${\bf f_{vv}}={\Phi_0^2d \over 8\pi^2\lambda^3}K_1(({\bf r}_i-{\bf r}_j)/\lambda) \protect\hat{\bf r}_{ij}$, 
  where $\protect\hat{\bf r}_{ij}=({\bf r}_i-{\bf r}_j)/|{\bf r}_i-{\bf r}_j|$.  
 Here the modified Bessel function $K_1$ is cut off beyond the distance $r=25\lambda$, where $\lambda$ is  the penetration depth (for Nb $\lambda=45$nm at $T=0$). The force $f$
  generated by the sawtooth pinning potential shown in Fig. \protect \ref{Fig1}
  is equal  to $f_1$ when $k\ell < x < k\ell+\ell_1$, and $f_2$ when
  $k\ell+\ell_1 < x < (k+1)\ell$, where $k=0,1,...,N-1$. We choose 
  $\ell_1=20\lambda=0.9\mu$m, $\ell_2=5\lambda=0.225\mu$m, $\ell=\ell_1+\ell_2$ and $N=10$, giving for the total width of the sample $w=11.25\mu$m. Its length (along  the $y$ direction) is set to $12\mu$m. The sample has
  periodic boundary conditions in both the $x$ and $y$ direction. 
  The Lorentz force due to the ac current is equal to $+f_{\rm L}$ for $T/2$
  time, and $-f_{\rm L}$ for $T/2$ using $T=0.3\mu$s. We considered the simplest case, in which the potential is induced
by thickness variations of a Nb superconductor thin film of thickness $d$, i.e. the SC
thickness, $d+h(x)$, changes along the $x$ direction, following a
sawtooth pattern. The pinning energy acting on the vortices is given
by $U(x)=(d+h(x))\epsilon_0$, where $\epsilon_0$ is the line energy of
the vortex per unit length. Thus the magnitudes of the forces acting on the vortices are
$f_1= \epsilon_0 \Delta h / \ell_1$ and $f_2=\epsilon_0 \Delta h / \ell_2$ 
for the two facets of the $\Delta h$ high teeth  (shown in Fig.\ \protect \ref{Fig1}c), and we choose $\Delta h = \ell_2$.
For Nb we have $\epsilon_0=1.7 \times 10^{-11}$N, the viscosity per unit length is $\eta_0=7\times 10^{-6}$Ns/m$^2$,
yielding $\eta=\eta_0d=1.4 \times 10^{-12}$Ns/m for a $d=2000${\AA} thick film. The total
  number of vortices in the simulation were $n=5$($\circ$),
  $n=250$($\triangleright$), and $n=500$($\diamond$) corresponding to 
  $\approx 0.7$G($\circ$), $35$G($\triangleright$), and $70$G($\diamond$) magnetic field in the sample.}
\label{Fig2}
\end{figure}

Since for high magnetic fields vortex-vortex interactions play an
important role, we have performed molecular dynamics simulations to
determine the ratchet velocity for a collection of vortices. As Fig.\ 
\ref{Fig2} demonstrates, we find that for low vortex densities the numerical results follow
closely the analytical prediction (\ref{vel1}), and the magnitude of the ratchet velocity decreases with increasing vortex density. The vortex densities used in the simulations correspond to an internal magnetic field of about 0.7, 35, and 70 G, covering a wide range of magnetic fields.
A key question for applications is if the ratchet velocity
(\ref{vel1}) is large enough to induce observable vortex motion at
experimentally relevant time scales. To address this issue in Fig.\ \ref{Fig2} we plotted
$v$ for Nb, a typical low temperature SC used in a wide range of
devices, for which the potential $U(x)$ is induced by {\it thickness variations} of the SC. The details of the model are described in the caption of Fig.\ \ref{Fig2}. As the figure indicates, the maximum ratchet velocity (5.2 m/s) is high enough to move a vortex across the typical few micrometer wide sample \cite{ruggiero} in a few microseconds. Furthermore, increasing the vortex density by two orders of magnitude decreases the vortex velocity only by a factor of three. 

\begin{figure}[htb]
\vskip 2.0 cm
\begin{flushleft}
{\bf (a)}
\end{flushleft}

\vskip -2.0 cm
\hskip 2.0 cm
\begin{center}
\mbox{\psfig{figure=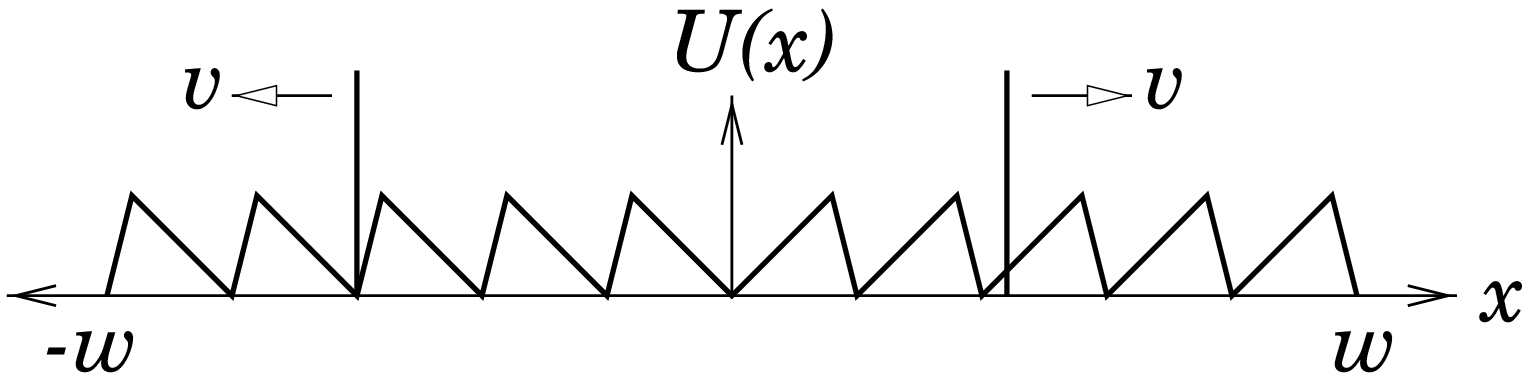,width=12 cm}}
\vskip 5.0 cm

\begin{flushleft}
{\bf (b)}
\end{flushleft}
\vskip -5.0 cm
\hskip 1.0 cm
\mbox{\psfig{figure=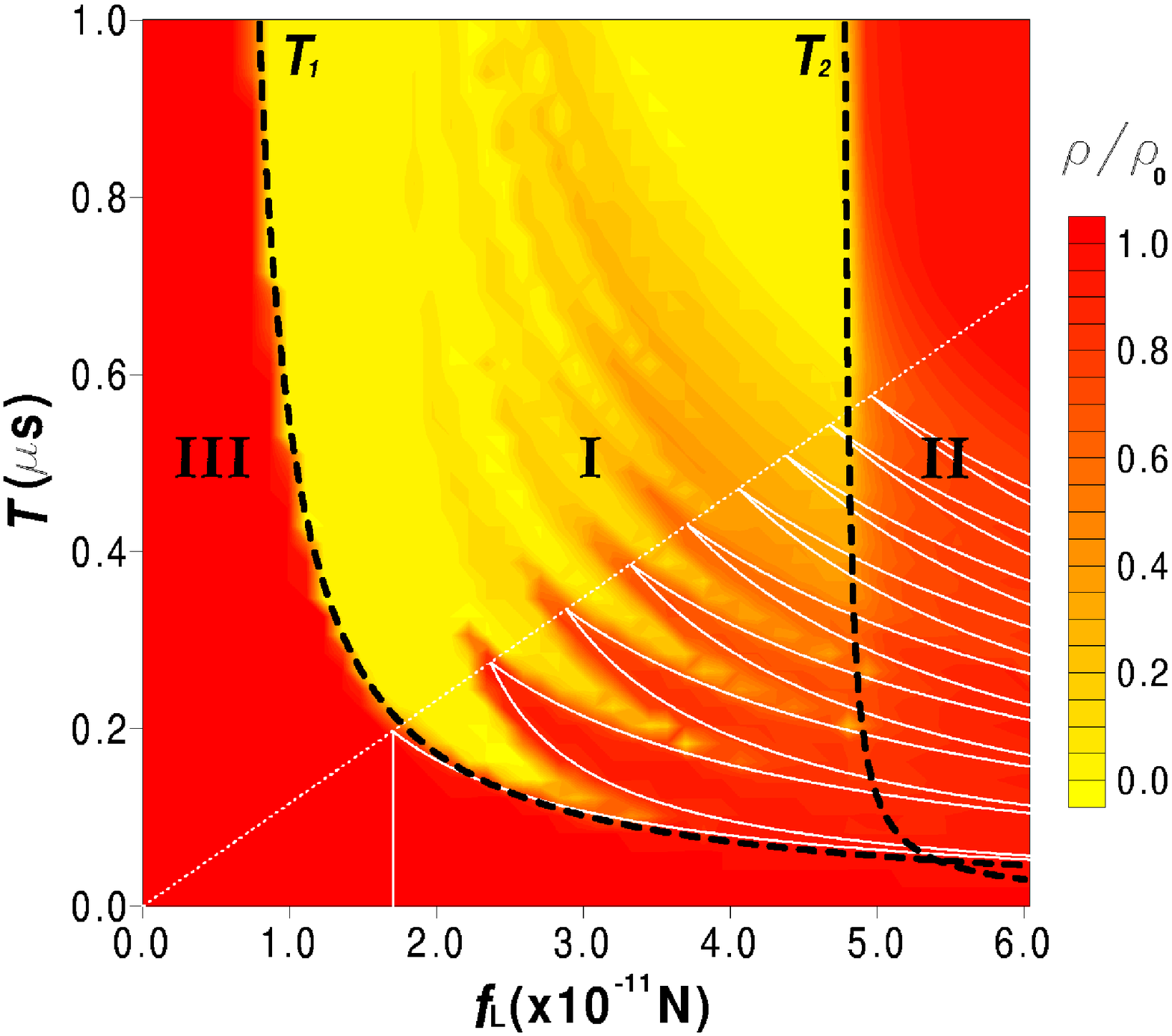,height=12 cm}}
\end{center}
\vskip -0.1 cm
\protect \caption{
Removing vortices from a SC using an asymmetric potential. {\bf (a)}
The potential necessary to remove vortices from the SC. Using the
simulation method described in Fig.\ \protect \ref{Fig2}, we investigated a
system consisting of $N=5$ teeth oriented to the left and the same
number oriented to the right, as shown in the figure, the parameters of
each tooth being identical to that described in Fig.\ \protect \ref{Fig1}c. To
mimic the pressure generated by the external magnetic field, which acts
to push vortices into the SC, on the two sides we attached two
reservoirs, that have a constant vortex density $\rho_0$ at all times.
Thus, vortices can leave the SC for the reservoir, or new vortices can
enter from the reservoir. In thin SC films, due to the Meissner
current, there is a geometrical barrier that acts to trap the vortices
inside the SC \protect \cite{zeldov1}. Since most applications of SC's involve
thin films, we included in the simulations this geometrical barrier,
that creates a force $f_{\rm in}(x)=-{{H\phi_0} \over {2 \pi}} x /
\sqrt{w^2-x^2}$ for $-w+d/2<x<w-d/2$, and $f_{\rm
edge}=2\epsilon_0-{{H\phi_0} \over {2 \pi}} \sqrt{4w/d-1}$ for $x>w-d/2$,
and $-f_{\rm edge}$ for $x<-w+d/2$. Thus the geometrical barrier opposes
the entry of the vortices at the edge of the SC, but once they move
inside, it moves them towards the center of the SC.  For successful
vortex removal the ratchet effect has to be strong enough to move the
vortices against $f_{\rm in}(x)$.  {\bf (b)} The $(f_{\rm L},T)$
diagram describing the effectiveness of the ratchet effect as a
function of the parameters characterizing the driving current, $f_{\rm
L}$. The color code corresponds to the the {\it relative vortex
density} $\rho/\rho_0$, where $\rho_0$ is the initial vortex density
corresponding to $H=1$G and $\rho$ is the final vortex density
after the application of the ac current. As the color code indicates,
there is a region where vortex removal is complete, the vortex density
being equal to zero. The dashed lines correspond to the $T_1$ and $T_2$
boundaries, that are calculated analytically (see text) and separate
the three main regimes: {\bf I}: complete vortex removal in the
majority of the regime, $\rho=0$; {\bf II}: partial vortex removal, $0
\leq \rho < \rho_0$; and {\bf III}: no change in the vortex density,
$\rho=\rho_0$. The thin white lines denote the boundaries of the regions where vortex trapping due to periodic orbits occurs. These boundaries correctly
reflect the structure of the fingers, but slightly deviate from the
results of the numerical simulation, because the
analytical calculation assumed an array of identical teeth.}
\label{Fig3}
\end{figure}

Next we discuss a potentially rather useful application of the ratchet
effect by demonstrating that it can be used to {\it drive vortices out
of a SC}. Consider a SC film that is patterned with two arrays of the
ratchet potential oriented in opposite directions, as shown in Fig.\ \ref{Fig3}a. During the application of the ac current the asymmetry of the potential in the
  right half moves the vortices in that region to the right, while vortices in the left half move to the left. Thus the vortices
  drift towards the closest edge of the sample, decreasing the vortex
  density in the bulk of the SC. We performed numerical simulations to
quantitatively characterize this effect, the details and the parameters being described in the caption of Fig.\ \ref{Fig3}. In Fig.\ \ref{Fig3}b we  summarize the effectiveness of vortex removal by plotting the reduced vortex density inside the SC as a function of the Lorentz force $f_{\rm L}$ and
the period $T$ of the current. One can see that there is a well  defined
region where the vortex density drops to zero inside the SC,
indicating that the vortices are completely removed from the
bulk of the SC. Outside this region we observe either a partial 
removal of the vortices or the ac current has no effect on the 
vortex density. 

The $(f_{\rm L},T)$ diagram shown in Fig.\ \ref{Fig3}b has three
major regimes separated by two boundaries. The $T_1=2\eta
\frac{\ell_1}{f_{\rm L}-[f_1+f_{\rm in}(-w+\ell_2)]}$ phase boundary
(here we assume $d/2<\ell_2$ and $f_{\rm in}(x)$ is defined in Fig.\ \ref{Fig3})
provides the time needed to move the
vortex all the way up on the $\ell_1$ long facet of the ratchet potential at the edge of the SC, i.e.  to
remove the vortex from the SC.
When $T<T_1$ the vortices cannot exit the SC.
The $T_2=2\eta \left( \frac{d/2}{f_{\rm L}-[f_2+f_{\rm edge}]} + \frac{\ell_2-d/2}{f_{\rm L}-[f_2-f_{\rm in}(-w+d/2)]} \right)$
phase boundary
(where $f_{\rm edge}$ is also defined in Fig.\ \ref{Fig3})
is the time needed for a vortex to enter from the edge
of the SC past the first potential maxima. Thus, when $T<T_2$ the
vortices cannot overcome the edge of the potential barrier.
These phase boundaries, calculated for non-interacting vortices, effectively determine the
vortex density in the three phases. Vortex removal is most effective in regime {\bf I}, where the vortices cannot move past the first potential barrier when they try to enter the SC, but they get past the barriers opposing their exit from the SC.
Thus the vortices are swept out of the SC by the ratchet effect, and no vortex can reenter, leading to a vortex density $\rho=0$.
Indeed, we find that the numerical simulations indicate complete vortex removal in the majority of this phase (see the contour lines in Fig.\ \ref{Fig3}b). An exception is the finger structure near the crossing of
the $T_1$ and $T_2$ boundaries. For fields and periods within the first finger
  (lowest in Fig.\ \ref{Fig3}b)  the vortex follows a periodic orbit inside a single potential well \cite{hanggi}. The subsequent fingers represent stable period orbits between two, three, or more wells, respectively. Since
the vortices cannot escape from these orbits, they remain trapped
inside the SC, increasing the vortex density within the fingers in the phase 
diagram. Fig.\ \ref{Fig3}b shows the analytically calculated
envelopes of the regions where such trapping occurs. An important feature of the finger structure is that
 stable periodic orbits, which prevent vortex removal, do not exist above the 
line $T_{\rm tip} = f_{\rm L} {2\eta\Delta U \over f_1 f_2 (f_2-f_1)}$ connecting 
the finger tips.  In regime {\bf II} vortices can enter the SC, but the ratchet effect is still sweeping them out, thus here we expect partial removal of the vortices, the final vortex density inside the SC being determined by the balance of vortex nucleation rate at the edge of the sample (which depends on the surface properties of the SC) and the ratchet velocity moving them out.
In regime {\bf III} the vortices cannot leave the SC and new vortices cannot enter the system, thus the initial density inside the SC is unchanged throughout this phase ($\rho=\rho_0$). 

Since the forces $f_{\rm in}(x)$ and $f_{\rm edge}$ depend on $H$, the
position of the phase boundaries $T_1$ and $T_2$ also depends on the
external magnetic field. In particular, there exists a critical field
$H^*$, such that for $H>H^*$ regime {\bf I}, where vortex removal is
complete, disappears, but regime {\bf II} with partial vortex removal
does survive. We find that for Nb films of geometry described in Fig.\
\ref{Fig3} we have $H^*\approx10$G. However, since $H^*$ is a consequence
of the geometric barrier, its value can be modified by changing the
aspect ratio of the film. Furthermore, for superconductors with
elliptic cross section the geometric barrier can be eliminated
\cite{clem}, thus phase {\bf I} with complete vortex removal could be
extended to high magnetic fields as well.

Vortex removal is important   for  numerous SC applications and  can
improve the functioning of  several devices.  An immediate application of the proposed method would be  improving the  operation of   superconducting  quantum
interference devices (SQUIDs), used as sensors in a wide assortment of
scientific    instruments \cite{ruggiero,muck,scott}. A
long-standing issue in    the performance of  SQUIDs is    $1/f$ noise
\cite{dantsker2,muck}, arising from the activated  hopping  of  trapped vortices \cite{tinkham}.  Reducing the
vortex density  in  these superconductors  is expected to   extend the
operation regime of these devices to lower frequencies.


Although over the past few years several applications of the ratchet
effect have been proposed, such as separating
particles
\cite{rousselet94,faucheux95}, designing molecular motors
\cite{kelly99}, smoothing surfaces
\cite{derenyi}, or rectifying the phase across a SQUID
\cite{hanggi1,hanggi2}, our proposal solves a well known acute problem of condensed matter physics, by {\it removing} vortices from a SC. In
contrast with most previous applications, which require the presence of
thermal noise, our model is completely deterministic.
Indeed, in Nb the variations in the pinning potential is
$\Delta U \approx 25$~eV, which is more than 10$^4$ times larger than
$k_{\rm B}T \approx 0.8$~meV at $T_c=9.26$~K, thus rendering thermal
fluctuations irrelevant. On the practical
side, a particularly attractive feature of the proposed method is that
it does not require sophisticated material processing to make it work:
First, it requires standard few-micron scale patterning techniques (the
micrometer tooth size was chosen so that a few teeth fit on a typical
SQUID, but larger feature size will also function if the period $T$ is
increased proportionally). Second, the application of an ac current
with appropriate period and intensity is rather easy to achieve.  For
applications where an ac current is not desired, the vortices can be
flushed out before the normal operation of the device. On the other
hand, if the superconducting device is driven by an ac current (e.g. rf
SQUIDs, ac magnets, or wires carrying ac current), the elimination of
the vortices will take place continuously during the operation of the
device. The analytically predicted phase boundaries, whose position is
determined by the geometry of the patterning, provide a useful tool for
designing the appropriate patterning to obtain the lowest possible
vortex density for current and frequency ranges desired for specific
applications. Finally, although here we limited ourselves to low
temperature SCs, the working principle of the ratchet effect applies to
high temperature superconductors as well.

\vskip 1.0cm


\vskip -0.4cm

{\bf Acknowledgements.} We wish to thank D. J. Bishop, S. N. Coppersmith, D. Grier, H. Jeong, A. Koshelev and S. T. Ruggiero for very useful discussions and help during the preparation of the manuscript. This research was supported by NSF Career Award DMR-9710998.

\vskip 0.3cm

Correspondence and requests for material should be addressed to A.-L.B. (e-mail:alb@nd.edu).
\end{multicols}

\newpage

\newpage

\newpage
\newpage


\normalsize


\begin{references}


\bibitem{tinkham} Tinkham, M. {\it Introduction to Superconductivity}
(McGraw-Hill, Inc., 1996).

\bibitem{trapped} Donaldson, G.B., Pegrum, C. M., \& Bain, R. J. P. 
  {\it Integrated Thin Film SQUID Instruments}, in {\it SQUID '85:
    Proceedings of the Third International Conference on
    Superconducting Quantum Devices, Berlin, June 25-28, 1995},
  edited by Hahlbohm H.D. \& L\"ubbig H. 749--753 (Walter Gruyter, Berlin-NY 1985).

\bibitem{ruggiero} Clarke, J. {\it SQUIDS: Principles, Noise and
    Applications} in {\it ``Superconducting Devices''}, edited by Ruggiero S.
  T. \& Rudman D. A. (Academic Press, 1990).

\bibitem{rpm} Blatter, G.,  Feigelman, M. V., Geshkenbeim, V. B., 
  Larkin, A. I. \& Vinokur, V. M. Vortices in High Temperature
  Superconductors. {\it Rev. Mod. Phys.} {\bf 66} 1125--1388 (1994).

\bibitem{koch} Koch, R.H., Sun,  J.Z., Foglietta, V. \& Gallagher, W.J.  
Flux dam, a method to reduce extra low-frequency noise when a
superconducting magnetometer is exposed to a magnetic field.
{\it Appl. Phys. Lett.}  {\bf 67}, 709--711 (1995).


\bibitem{dantsker2} Dantsker, E., Tanaka, S. \& Clarke, J. High-$T_c$ Superconducting Quantum Interference Devices with
    Slots or Holes: Low $1/f$ Noise in Ambient Magnetic Fields.
  {\it Appl. Phys. Lett.} {\bf 70}, 2037--2039 (1997).


\bibitem{magnasco} Magnasco, M. O. Forced Thermal Ratchet.
  {\it Phys. Rev. Lett.} {\bf 71}, 1477--1481 (1993).

\bibitem{astumian} Astumian, R. D. Thermodynamics and Kinetics of
    a Brownian Motor. {\it Science} {\bf 276}, 917--922 (1997).

\bibitem{julicher} J\"ulicher, F., Ajdari, A. \& Prost, J. 
  Modeling Molecular Motors. {\it Rev. of Mod. Phys.} {\bf 69},
  1269--1281 (1997).

\bibitem{hanggi} H\"anggi, P. \& Bartussek, R. Brownian Rectifiers: How to Convert Brownian Motion into Directed Transport. 
in {\it Lecture Notes in Physics Vol. 476, pp. 294-308}, edited by J. Parisi et al. (Springer, Berlin, 1996).

\bibitem{clem} Clem, J. R., Huebener, R. P. \& Gallus, D. E. Gibbs Free-Energy Barrier Against Irreversible Magnetic Flux Entry into a Superconductor. {\it J. Low Temp. Phys.} {\bf 12}, 449--477 (1973).

\bibitem{muck} M\"uck, M. Practical Aspects for SQUID
    Applications. {\it Superlattices and Microstructures} {\bf 21}, 415--421 (1997).



\bibitem{scott} Scott, B. A., Kirtley, J. R., Walker, D., Chen, B.-H. \&
  Wang, Y.  Application of Scanning SQUID Petrology to
    High-Pressure Materials Science. {\it Nature} {\bf 389}, 164--167 (1997)

\bibitem{rousselet94} Rousselet, J., Salome, L., Ajdari, A. \& Prost, J.
Directional motion of brownian particles induced by a periodic
asymmetric potential.
{\it Nature} {\bf 370}, 446-448 (1994).

\bibitem{faucheux95} Faucheux, L. P., Bourdieu, L. S., Kaplan, P. D. \& Libchaber, A. J. Optical thermal ratchet. {\it Phys. Rev. Lett.} {\bf 74}, 1504-1507 (1995).

\bibitem{kelly99} Kelly, T. R., Silva, H. D. \& Silva, R. A.
 A rationally designed molecular motor.
{\it Nature} (submitted).

\bibitem{derenyi} Der\'enyi, I, Lee, C.-S. \& Barab\'asi, A.-L.  Ratchet effect in surface electromigration: smoothing surfaces by an AC field. {\it Phys. Rev. Lett.} {\bf 80}, 1473--1476.

\bibitem{hanggi1} Zapata, I., Bartussek, R., Sols, F. \& H\"anggi, P. Voltage Rectification by a SQUID Ratchet. {\it Phys. Rev. Lett.} {\bf 77}, 2292--2295 (1996).

\bibitem{hanggi2} Zapata, I., Luczka, J., Sols, F. \& Ha\"nggi, P.  Tunneling Center as a Source of Voltage Rectification in Josephson Junctions. {\it Phys. Rev. Lett.} {\bf 80}, 829--832 (1998).

\bibitem{Nori} Reichhardt, C., Olson, C.J. \& Nori, F. 
  Nonequilibrium dynamic phases and plastic flow of driven vortex lattices in superconductors with periodic arrays of pinning sites. {\it Phys. Rev. B} {\bf 58} 6534--6564 (1998).

\bibitem{reichhardt78} Reichhardt, C., Olson, C.J. \& Nori, F. Dynamic phases of
    vortices in superconductors with periodic pinning.
    {\it Phys. Rev. Lett.} {\bf 78}, 2648--2651 (1997).

\bibitem{olson98} Olson, C. J., Reichhardt, C. \& Nori, F. Nonequilibrium Dynamic Phase Diagram for Vortex Lattices. {\it Phys. Rev. Lett.} {\bf 81}, 3757--3760 (1998). 

\bibitem{zeldov1} Zeldov, E. {\it et al.} Geometrical barriers in high-temperature superconductors. {\it Phys. Rev. Lett.} {\bf 73}, 1428--1431 (1994).

\end{references}
\end{document}